\documentclass{aa}  
\usepackage{natbib}
\usepackage{graphicx}
\usepackage[squaren,Gray]{SIunits}
\usepackage{color, colortbl}
\usepackage{txfonts}

\begin{document} 

\title{Very deep images of the innermost regions of the $\beta$ Pictoris debris disc at $L^\prime$}

   \author{
   	J. Milli   \inst{1,2}
	\and	A.-M. Lagrange \inst{1} 
          \and D. Mawet \inst{2}
	\and O. Absil \inst{3}
	\and J.-C. Augereau \inst{1}
          \and D. Mouillet   \inst{1}
          \and A. Boccaletti \inst{4}
          \and  J. H. Girard \inst{2}         
         	\and G. Chauvin \inst{1}
          }

   \institute{Universit\'e Grenoble Alpes, IPAG, F-38000 Grenoble, France \\
CNRS, IPAG, F-38000 Grenoble, France 
   		\and
  		European Southern Observatory (ESO), Alonso de C\'ordova 3107, Vitacura, Casilla 19001, Santiago, Chile
   		\and
  		D\'epartement d`Astrophysique, G\'eophysique et Oc\'eanographie (AGO), Universit\'e de Li\`ege, 17 All\'ee du Six Ao\^ut, B-4000 Li\`ege, Belgium  
		\and 
		 LESIA, Observatoire de Paris, CNRS, Universit\'e Pierre et Marie Curie 6 and Universit\'e Denis Diderot Paris 7, 5 place Jules Janssen, 92195 Meudon, France
             }

   \date{Received 26 November 2013 / Accepted 24 April 2014}
   
  \abstract
{Very few debris discs have been imaged in scattered light at wavelengths beyond $\unit{3}{\micro\meter}$ because the thermal emission from both the sky and the telescope is generally too strong with respect to the faint emission of a debris disc. We present here the first analysis of a high angular resolution image of the  disc of $\beta$ Pictoris at $\unit{3.8}{\micro\meter}$. }
{Our primary objective is to probe the innermost parts of the $\beta$ Pictoris debris disc and describe its morphology. We performed extensive forward modelling to correct for the biases induced by angular differential imaging on extended objects and derive the physical parameters of the disc.}
{This work relies on a new analysis of seven archival datasets of $\beta$ Pictoris observed with the NaCo instrument at the Very Large Telescope in the $L^\prime$ band, including observations made with the Annular Groove Phase Mask vortex coronagraph in 2013. The data analysis consists of Angular Differential Imaging associated with disc forward modelling to correct for the biases induced by that technique. The disc model is subtracted from the data and the reduction performed again in order to minimize the residuals in the final image.}
{The disc is detected above a $5\sigma$ level between 0.4\arcsecond and 3.8\arcsecond. The two extensions have a similar brightness within error bars. We confirm an asymmetry previously observed at larger distances from the star and at shorter wavelengths: the isophotes are more widely spaced on the north-west side (above the disc apparent midplane) than on the south-east side. This is interpreted as a small inclination of the disc combined with anisotropic scattering. Our best-fit model has an inclination of $86^\circ$ with an anisotropic Henyey-Greenstein coefficient of $0.36$.
This interpretation is supported by a new asymmetry detected in the disc: the disc is significantly bowed towards the north-west within $3\arcsec$ (above the apparent midplane). We also detect a possible new asymmetry within $1\arcsec$, but at this stage we cannot discern between a real feature and an underlying speckle.}
 {}
  

   \maketitle
%

\section{Introduction}
\label{sec_intro}
The debris disc around $\beta$ Pictoris is one of the most studied planetary systems. \citet{Smith1984} discovered the debris disc when the Infrared Astronomical Satellite (IRAS) detected an infrared excess associated with the star \citep{Aumann1985}, and since then it has been studied from the visible and ultraviolet to the millimeter wavelengths. In optical/near infrared scattered light, the debris disc appears as a  bright dust disc seen almost edge-on \citep[e.g.][]{Kalas1995}. It can be decomposed into two components \citep{Augereau2001}: an inner warped disc detected up to around 150AU and an outer main disc extending up to 1000AU \citep{Smith1987}. 
\citet{Currie2013} reported the detection of the disc with the NAOS-CONICA (NaCo) instrument at the Very Large Telescope (VLT) in the $L^\prime$-band,  but did not investigate and discuss its properties. Observations in this wavelength range still dominated by scattered light  from the star, but close to the transition with thermal emission, could bring precious additional constraints on the dust properties. In addition, adaptive optics (AO) corrections from ground-based telescopes perform better at longer wavelengths, with Strehl ratios typically between 60\% and 80\%, while the angular resolution ($\sim0.1\arcsec$) is still sufficient to probe the disc inner regions, typically within a few tens of astronomical units. These regions are critical because the disc interacts with potential planets therein. For instance, \citet{Mouillet1997} attributed the observed inner warp of the disc to the gravitational perturbation of a massive body located on an inclined orbit. This perturber, discovered later by \citet{Lagrange2009}, was indeed a giant planet orbiting within 10 AU from the star \citep{Chauvin2012}, and the inclination of its orbit with respect to the main outer disc was confirmed \citep{Lagrange2012}. The position angle (hereafter PA) of the main disc was found to be $29.0^\circ \pm{0.2^\circ}$, whereas the inner warp is offset by ${3.9^\circ}^{+0.6^\circ}_{-0.1^\circ}$.  

Imaging debris discs in the  $L^\prime$-band centred around $\unit{3.8}{\micro\meter}$ is a difficult task. The thermal emission from the telescope/instrument and the sky emission, called hereafter the background emission, is much larger than at shorter wavelengths, while the star to disc contrast remains the same\footnote{assuming a constant dust albedo}. Quasi-static speckles from imperfect optics prevail inside 1\arcsecond whereas background emission is the dominant noise source beyond that separation. Therefore, high contrast and high angular resolution techniques are required, as well as specific data reduction tools. Innovative methods such as angular differential imaging \citep[ADI,][]{Marois2006}, locally-optimized combination of images \citep[LOCI,][]{Lafreniere2007}, and principal components analysis \citep[PCA,][]{Soummer2012,Amara2012} were developed to detect and characterize faint companions. They are well suited to image  point sources. For discs, they are also used to reveal high spatial frequency features (\citealt{Buenzli2010}; \citealt{Thalmann2011}; \citealt{Boccaletti2012}; \citealt{Lagrange2012} or \citealt{Lagrange2012_HR4796}), but any extended structure is strongly affected by these image processing techniques \citep{Milli2012,Esposito2014}. In this paper, we present new images of the disc obtained with VLT/NaCo in the $L^\prime$-band (section \ref{sec_obs}). We detail the morphology from $3.8\arcsec$ down to 0.4\arcsecond (section \ref{sec_charac}), and perform forward modelling to correct for the biases induced by ADI (section \ref{sec_interpretation}). Finally, we discuss our results in section \ref{sec_discussion}, before concluding in section \ref{sec_conclusion}.

\section{Observations and data reduction}
\label{sec_obs}
\subsection{Observations}

\begin{table}
\caption{Description of the seven data sets}             
\label{tab_dataset}      
\centering                          

\begin{tabular}{ cccccc}      
\hline\hline                        
Date & $\Delta \theta$\tablefootmark{a}  & $\Delta t$\tablefootmark{b} (hrs) &  Seeing & $\tau_0$ (ms)   & Sr\tablefootmark{c} (\%)  \\    
\hline                        
26/12/09 &  $44^\circ$ & 1.3 & 0.6\arcsec & 8 & 83 \\ 
28/09/10 &  $51.5^\circ$  &  1.6 & 0.9\arcsec & 1.9  & 56 \\        
17/11/10 & $54.3^\circ$ & 1.2 & 1.4\arcsec & 1.4 & NA \\ 
12/10/11 &  $54.1^\circ$ & 1.3 &  0.65\arcsec & 5 & 62 \\       
11/12/11 & $54.1^\circ$ & 0.8 & 1.4\arcsec & 4.4 & 63\\ 
16/12/12 & $67.6^\circ$ & 0.9 & 0.75\arcsec & 5.5 & 70 \\ 
31/01/13 & $84^\circ$ & 1.6 & 1.0\arcsec & 2.2 & 75 \\ 
\end{tabular}
\tablefoot{
\tablefoottext{a}{\small{Parallactic angle variation.}}
\tablefoottext{b}{\small{Effective on-source integration time.}}
\tablefoottext{c}{\small{Strehl ratio.}}
}
\end{table}
   
The star was observed on 31 January 2013 during the science verification observing run of NaCo's Annular Groove Phase Mask (AGPM) coronagraphic mode \citep{Absil2013}.  The AGPM coronagraph operates in the $L^\prime$-band at $\unit{3.8}{\micro\meter}$ with an undersized Lyot stop blocking the light from the secondary mirror and telescope spiders \citep{Mawet2013}. Because the signal-to-noise ratio (SNR) was not high enough on this individual data set to retrieve the disc morphology with sufficient accuracy\footnote{The AGPM coronagraphic observations with NaCo at the VLT showed a slightly degraded sensitivity in the background compared to saturated imaging, because of the lower throughput of the chosen NaCo Lyot stop, and the less efficient nodding scheme inherent to coronagraphy (dithering is not possible).}, we have used six additional data sets that had been recorded for orbit monitoring. These additional data were obtained in pupil-tracking mode, the star being saturated. A summary of the seven data sets with the average observing conditions is shown in Table \ref{tab_dataset}.

\subsection{Data reduction of individual datasets}
\label{sec_data_red}

For the AGPM observations, the sky subtraction was different from that described in \citet{Absil2013}. The main source of noise that needs to be overcome to reveal faint extended emission beyond $1\arcsecond$ comes from the high and variable background at this wavelength. The standard way to evaluate the background in high contrast coronagraphic imaging is to open the AO loop and perform an offset of the telescope to measure an nearby empty sky region. This technique does not provide a sufficient accuracy in our case because the telescope thermal emission significantly changed, depending on the AO loop status (open or closed). Therefore, the cosmetic treatment of the raw data only consisted of dark subtraction, flat-field correction, and bad-pixel interpolation. The background is removed in a later stage, during the ADI subtraction process, along with the stellar halo.  The data were binned every 100 seconds, resulting in a data cube of 90 images, and the images were re-centred with the same procedure as described in  \citet{Absil2013}, leading to an uncertainty in the AGPM centring of 0.1 mas and an uncertainty in the star centring with respect to the AGPM centre of 8.5 mas.

In the case of the six non-coronagraphic data sets, the sky was evaluated using a different telescope offset position and subtracted from the raw images. The same cosmetic treatment was then applied. Because of the small telescope offsets, the total field of view is limited to a radius of $3.8\arcsec$. The star centre position is obtained from a fit to the low-flux level wings of the saturated image and has an uncertainty of 7 mas, obtained with the same method described in the appendix of \citet{Lagrange2012}.

Then, the star subtraction algorithm consisted of a PCA performed in  two to five concentric annuli between $0\arcsecond$ and $3.8\arcsecond$. This method uses a truncated basis of eigenvectors created by a Karhunen-Lo\`eve transform of the initial set of images, to perform the subtraction of the point-spread function, hereafter PSF. The number of principal components retained varied for each data set between 1 and 4. It was adjusted visually to maximize the disc SNR on each individual reduction. We did not do any frame selection to maximize the exposure time, thus the signal from the disc .

\subsection{Combination of the data and subtraction of $\beta$ Pictoris b}

\begin{table}
\caption{Calibration of the data sets and properties of $\beta$ Pictoris b}             
\label{tab_prop_bpicb}      
\centering                          
\begin{tabular}{ p{0.15\linewidth}  p{0.1\linewidth} p{0.15\linewidth} c p{0.15\linewidth} }        
\hline \hline 
Date & True north & Separation (mas) & PA & Contrast (mag) \\    
\hline                        
26/12/09 & $-0.06^\circ$ & $299^{+13}_{-13}$ & ${211.1^\circ}^{+1.7}_{-1.6}$ & $7.7^{+0.25}_{-0.20}$ \\ 
28/09/10 & $-0.36^\circ$ & $385^{+10}_{-11}$ & ${210.1^\circ}^{+1.5}_{-1.5}$ & $8.1^{+0.26}_{-0.21}$ \\        
17/11/10 & $-0.25^\circ$ & $392^{+13}_{-14}$ & ${211.4^\circ}^{+1.8}_{-1.7}$ & $7.7^{+0.26}_{-0.21}$ \\ 
12/10/11 & $-0.35^\circ$  \tablefootmark{a}  & $439^{+4}_{-5}$ & ${212.9^\circ}^{+0.5}_{-0.5}$ & $7.9^{+0.10}_{-0.09}$ \\       
11/12/11 &  $-0.35^\circ$ & $441^{+3}_{-3}$ & ${212.9^\circ}^{+0.3}_{-0.3}$ & $8.1^{+0.07}_{-0.06}$ \\ 
16/12/12 &  $-0.51^\circ$ & $449^{+6}_{-5}$ & ${211.6^\circ}^{+0.7}_{-0.6}$ & $8.0^{+0.07}_{-0.06}$ \\ 
31/01/13 &  $-0.45^\circ$ & $448^{+3}_{-4}$ & ${212.3^\circ}^{+0.3}_{-0.3}$ & $8.0^{+0.11}_{-0.11}$ \\ 
\end{tabular}
\tablefoot{
\tablefoottext{a}{\small{No information available for this date, we assumed the same value as for 11/12/11}.}
}
\end{table}

On the individual data sets, the residual thermal noise is non-gaussian at $L^\prime$ and shows large patterns of the size of several resolution elements. 
For instance, for resolution elements between $0.5\arcsecond$ and $1\arcsecond$, the Shapiro-Wilk test (Shapiro \& Wilk 1965) yields a p-value below $2\%$ for all individual data sets, except for the AGPM image (p-value of $90\%$), and this is highly dependent on the number of principal components retained for each data set. In order to increase the disc SNR, we combined several uncorrelated data taken at different epochs. This represents the first approach of that kind in high contrast imaging. Combining seven data sets reduces the standard deviation by a factor 2.7 for the resolution elements between $1\arcsecond$ and $2\arcsecond$, which is also what is  expected from the central limit theorem assuming uncorrelated noise ($\sqrt{7}=2.64$).


For each data set, an AO-corrected unsaturated PSF recorded before and/or after the science observations was used as a photometric calibrator. The six final images were individually scaled to the total flux of the central star $\beta$ Pictoris, cropped to the same field of view of radius $3.8\arcsecond$, and averaged. The total flux of the star was calculated by performing aperture photometry on the AO-corrected unsaturated PSF. The flux increases with the aperture radius until convergence is reached for a radius of $1.35\arcsecond$, therefore, we used this radius to make sure we encircled the total star flux and are not subject to the PSF variability due to the AO correction. Note that the disc is well below the noise level in the unsaturated PSF and does not bias the photometry of the star.

\begin{figure}
\includegraphics[width=0.5\textwidth]{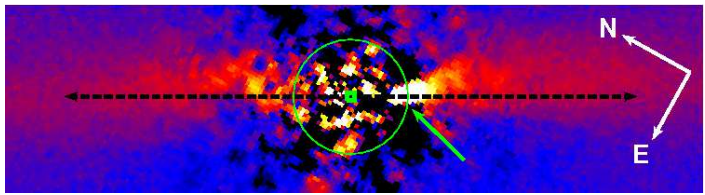}
\caption{PCA-reduced image obtained after combination of the seven individual images, colour scale-adjusted to show the smearing of the planet (green arrow), which prevents disc signal extraction at this location. The green circle has a radius of $0.4\arcsecond$ and is centred on the star (green square).  The black arrow has a position angle of 29$^\circ$ and a total length of $6\arcsec$.}
\label{fig_smearing_planet}
\end{figure}

As visible in Fig. \ref{fig_smearing_planet}, the presence of $\beta$ Pictoris b prevents from detecting the disc inside $0.6\arcsec$ on the south-west (SW) extension. The planet orbital motion between 2009 and 2013 induces a bright dot smeared along 5.5 px, or 150 mas. To disentangle any disc structure or additional companions from the signature of $\beta$ Pictoris b, we subtracted the planet signal from the seven data sets. We used the negative fake planet technique described in \citet{Lagrange2010} and \citet{Bonnefoy2011} to retrieve the photometry and astrometry of the planet at each epoch. We used the same optimization area as described in \citet{Absil2013}. The photometry and astrometry derived for the position of the planet in each individual data set are listed in Table \ref{tab_prop_bpicb}. The astrometry error budget includes the statistical error in the companion position calculated with negative fake planets as in \citet{Absil2013}, the calibration errors from the platescale, true north and rotator offset, and the error in the star centre described in section \ref{sec_data_red}. The photometry error budget includes the statistical error plus a contribution accounting for photometric variations of 0.05 mag for a photometric night (ESO data quality control standards). 

\begin{figure}
\includegraphics[width=0.5\textwidth]{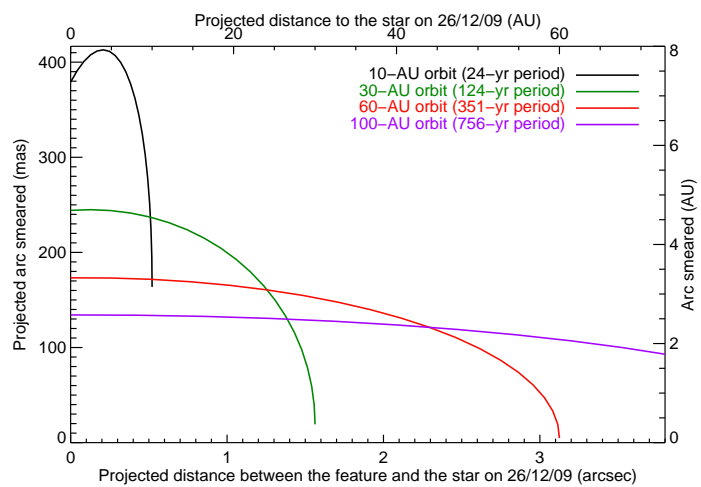}
\caption{Projected distance smeared by a potential disc feature or planet in keplerian motion around the star during the time span between the first and last observations (3.1yrs).}
\label{fig_smearing_disc_feature}
\end{figure}

Similarly, disc non-axisymmetric features or additional point sources in orbital motion around the star would also smear an arc in the image. Assuming these features are in keplerian motion around the star, we explored the amplitude of this effect and show the results in Fig. \ref{fig_smearing_disc_feature} for four different semi-major axis. A feature orbiting at 30AU from the star can smear an arc of 240 mas, and this value goes down to 170 mas for a 60AU-orbit and 135 mas for a 100AU-orbit. Therefore, we cannot exclude that our image contains disc features smeared over a few hundred mas that would otherwise have appeared more compact.

\section{Characterization of the disc}
\label{sec_charac}

\subsection{Correction for biases due to ADI}
\label{sec_bias_correction}

The signal of extended objects may be heavily influenced by ADI \citep{Milli2012}. Therefore, a careful calibration was performed. We used three approaches, including: 
\begin{enumerate}
\item iterations as in \citet{Lagrange2012} in order to obtain images that have the least biases from ADI;
\item forward modelling to constrain the dust density distribution as in \citet{Thalmann2014}, presented in  section \ref{sec_single_pop}; and
\item injection of model discs in the data to calibrate the morphological parameters measured in the images, as in \citet{Boccaletti2012}. This is presented in section \ref{sec_discussion}.
\end{enumerate}

\begin{figure*}
\centering                          
\includegraphics[width=0.8\textwidth]{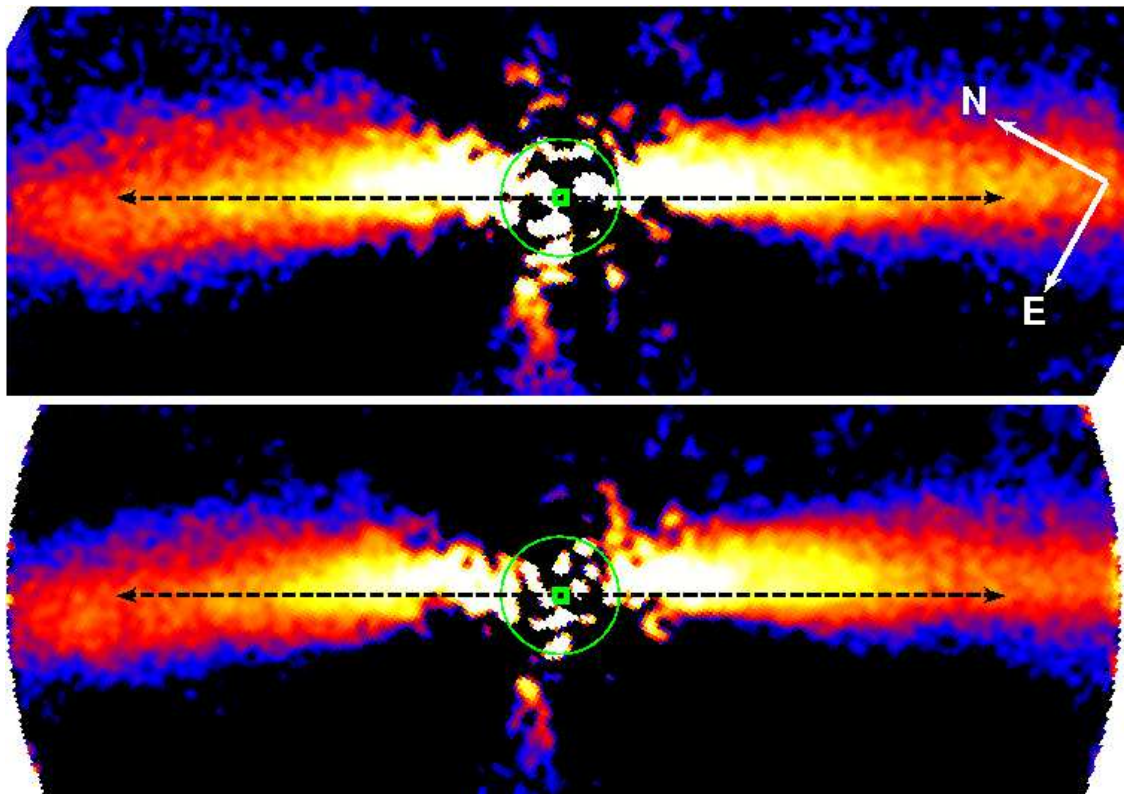}
\caption{Two different disc reductions: cADI-disc (top) and PCA-disc (bottom). The black arrow shows the apparent midplane, has a position angle of $29^\circ$ (equal to that of the main disc), and a total length of 6\arcsec. The colour scale is identical for both images but not linear (square root). The green square marks the position of the star and the green circle has a radius of $0.4\arcsec$.}
\label{fig_disc_cadi_pca}
\end{figure*}

The iterative approach was already implemented in \citet{Lagrange2012} and is called cADI-disc, where cADI refers to classical ADI and consists of a simple subtraction of the median of the cube of images to subtract the star. The resulting image is shown in Figure \ref{fig_disc_cadi_pca} (top). Here, we also adapted this technique for the PCA, namely we ran a first reduction to get an image of the disc. After applying a mask to this image to set the regions unaffected by the disc to 0, we subtracted this image from the initial data cube. We therefore obtained a library that ideally does not contain any astrophysical signal and we applied a PCA on our initial data cube using this reference library. The reduced image is shown in Fig. \ref{fig_disc_cadi_pca} (bottom). The disc in the cADI-disc image is slightly thicker than in the PCA-disc image because disc self-subtraction is higher with the latter algorithm. We will therefore use the cADI-disc image to extract the surface brightness profiles of the disc. Speckle noise is weaker in the PCA-disc image, so we will use this image to extract the morphological parameters of the disc other than the vertical thickness. It should be noted that this iterative technique minimizes disc self-subtraction but it does not totally prevent it, in particular, in faint vertically extended parts of the disc, far from the midplane. The pixels of the image that contain the disc signal at all observed parallactic angles (because the disc is vertically too extended) still suffer from disc self-subtraction. This affects essentially the pixels inside $0.8\arcsec$ and is very clearly visible on the PCA-reduced image of a disc model (Fig. \ref{fig_disc_subtraction}, middle, described later in section \ref{sec_result_single_comp}). Space-based observations that do not use field rotation for PSF subtraction indeed reveal a faint emission at all azimuths in the sky plane at a radius of $0.5\arcsec$ (e.g. Fig. 2 of \citealt{Golimowski2006}).

\subsection{Disc morphology}
\label{sec_disc_morphology}

\begin{figure}
\includegraphics[width=0.5\textwidth]{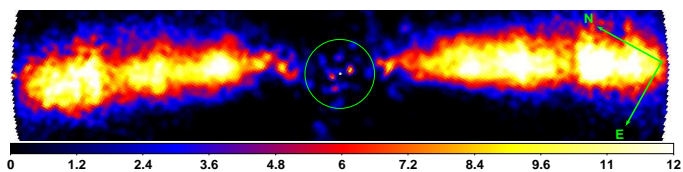}
\caption{SNR map of the PCA-disc image. The colour scale is linear and spans from 0 to 10. The green circle indicates  the start of the $5\sigma$ detection at a radius of $0.4\arcsec$.}
\label{fig_SNR_map}
\end{figure}

The disc is detected unambiguously from underlying speckle and thermal noise up to the maximum field of view available, i.e. $3.8\arcsec$, and down to $0.4\arcsec$. This is the closest separation ever reached on the disc. The SNR map of the disc is shown in Fig. \ref{fig_SNR_map}. In the subsequent sections, we use the terminology spine and apparent midplane of the disc to refer to the curve joining the brightest pixels of the disc in each vertical profile, and to the line going through the star with a PA of $29^\circ$, corresponding to the best estimate of the PA of the main outer disc at Ks, as seen in section \ref{sec_intro}.

\subsubsection{Offset and bow of the disc}

\begin{figure*}
\centering                          
\includegraphics[width=0.8\textwidth]{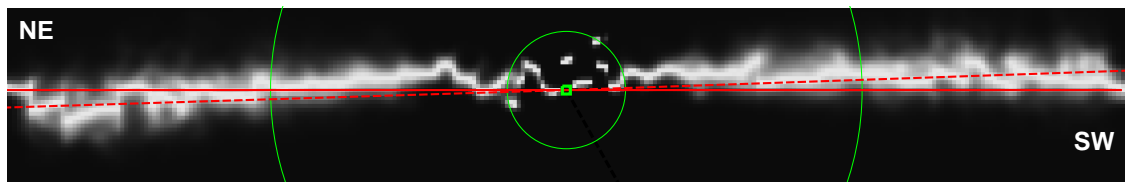}
\caption{PCA-disc image where each vertical profile has been normalized by the maximum spine brightness to enhance the vertical position of the spine. The two green circles have a radius of $0.4\arcsec$ and $2\arcsec$ respectively. The PA of the main disc ($29^\circ$) as measured by \citet{Lagrange2012} is indicated by a solid red line, whereas the best fit PA of $30.8^\circ$ (NE) and $211.0^\circ$ (SW) as measured from the $L^\prime$ images are shown with dashed lines. The waviness of the spine is discussed in section \ref{sec_ripples}}
\label{fig_spine_position}
\end{figure*}

\begin{figure}
\includegraphics[width=0.5\textwidth]{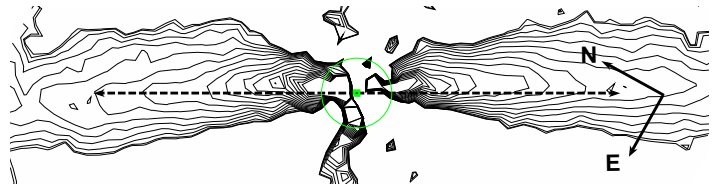}
\caption{Isophotes corresponding to the cADI-disc image of Fig. \ref{fig_disc_cadi_pca}. The black arrow passes through the star (green square), has a position angle of $29^\circ$, and a total length of $6\arcsec$.}
\label{fig_disc_cadi_isophotes}
\end{figure}

The disc is clearly vertically offset from the star within $2\arcsec$ : the star lies below the spine. This is illustrated in Fig. \ref{fig_spine_position}, in which each vertical profile has been normalized to its peak value to highlight the position of the spine. At $1\arcsec$, the spine is about 2AU above the midplane of the disc. This is also clearly visible on the isophotes of the disc in Fig. \ref{fig_disc_cadi_isophotes}.

Moreover, these two figures show that the disc is bowed towards the north-west (NW), i.e. above the apparent midplane, between $-3\arcsec$ and $3\arcsec$. This bow of the spine in the innermost regions was already marginally detected by \citet{Boccaletti2009} in the H and Ks band, and was reported in the optical by \citet{Golimowski2006} who mentioned it for the SW extension. This bow can be reproduced by anisotropic scattering of the dust, combined with a slight inclination of the disc with respect to an edge-on view, and it will be discussed later.
Two additional observations support this interpretation. First, the isophotes are more widely spaced above the apparent midplane (NW side) than below (SE side). This was previously observed by \citet{Golimowski2006}. Second, the disc appears thicker above the spine than below it (see Fig. \ref{fig_disc_cadi_pca} and \ref{fig_disc_cadi_isophotes}).

\subsubsection{Position angle of the disc}

We extracted the PA of the disc along with the position and maximum brightness of the spine from our cADI-disc and PCA-disc images with the same technique as described in \citet{Lagrange2012} for the Ks image. In the Ks band, \citet{Lagrange2012} found a PA of  $29.3^\circ \pm{0.2^\circ}$ for the main disc by fitting a Lorentzian profile beyond $4.8 \arcsecond$. To measure the PA of the warped inner component, \citet{Lagrange2012} decomposed the vertical profiles into two Lorentzian profiles and found in that case a PA of  $29.0^\circ \pm{0.2^\circ}$ and ${32.9^\circ}^{+0.6^\circ}_{-0.1^\circ}$ for the main and warped discs (see Table 2 in \citealt{Lagrange2012}).

Unlike in the Ks image, the field of view available in our $L^\prime$ image is too small to access a region dominated only by the main disc, which would allow us to determine the PA of  this outer component and derotate the images. If we try to find the best PA of the disc by nulling the spine vertical position on average between $3\arcsec$ and $3.8\arcsec$, we get a PA value for the north-east (NE) and SW extensions of $30.8^\circ  \pm{0.6^\circ} $ and $211.0^\circ  \pm{0.7^\circ}$. The same technique applied to our Ks images yields a similar PA within error bars ($30.2^\circ  \pm{0.5^\circ} $ and $210.5^\circ  \pm{0.6^\circ}$), confirming that the $L^\prime$ image is compatible with the Ks image and confirming that our view of the disc is a superimposition of the two contributions of the outer main disc and the inner warped disc. Fig. \ref{fig_spine_position} shows the lines of PA $29^\circ/209^\circ$ and $30.8^\circ/211.0^\circ$ on top of the spine location. This shows that within $3.5\arcsec$ the optimal PA of $30.8^\circ/211.0^\circ$ (red dashed line) is indeed the most compatible with the L' data. At the largest separations beyond $3.5\arcsec$, the disc tends to align again on the PA of the main disc (red solid line). We show in section \ref{sec_double_pop} that this is perfectly compatible with a two-component disc if the inner component is brighter than the outer component within $3.5\arcsec$.

The dynamical range is not high enough, however, to allow us to fit a two-component Lorentzian profile and separate the contributions of the main disc from that of the warped disc. Therefore we decided to use the PA of $29^\circ$ measured from the Ks image as an a priori to derotate our L' image.

\subsubsection{Position of the spine}

\begin{figure}
\includegraphics[width=0.5\textwidth]{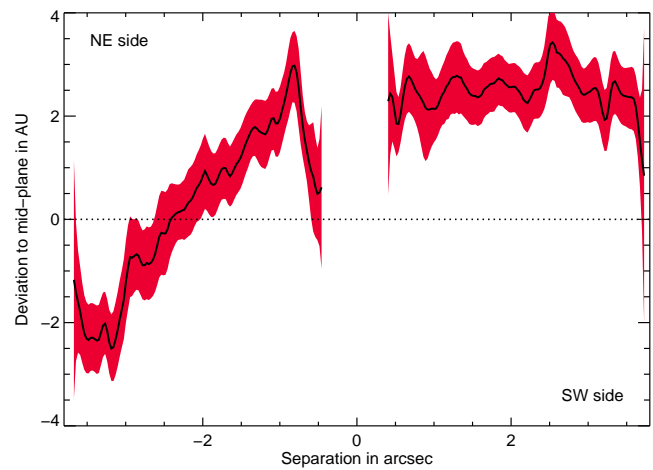}
\caption{Departure of the spine with respect to the apparent midplane (defined in section \ref{sec_disc_morphology} as the plane at $29^\circ$ PA) as measured on the PCA-disc image. NE is to the left. The uncertainty, shown here in red, is the $3\sigma$ error in the centroid position of the Lorentzian profile fitted to each vertical profile, after setting a radially-dependent noise level to each pixel. The waviness of the curve is due to the background and remaining speckle noise, as shown in section \ref{sec_ripples}. We note, however, a  larger ripple at $0.8\arcsec$ on the NE side.}
\label{fig_disc_center}
\end{figure}

The spine position as measured from the lorentzian fit is displayed in Fig. \ref{fig_disc_center}. The bow of the disc is clearly confirmed by this plot and its extension can be determined: the spine is a concave function of the separation between $-3.\arcsec$ and $3.5\arcsec$. On the NE side, the spine of the disc goes below 0 beyond $2.5\arcsec$ and the slope changes sign beyond $3\arcsec$. This fact is also well reproduced by the model of the disc presented in section \ref{sec_double_pop}. The spine of the disc versus separation is not a smooth curve but present ripples. We show in section \ref{sec_ripples} that these small-amplitude ripples are due to the background and speckle noise and a smooth curve is very likely, given the error bars, however, we note a larger ripple at $0.8\arcsec$ on the NE extension, which is very clear in Fig. \ref{fig_spine_position}. This deviation is at $2\sigma$ above the noise level and its significance will be discussed in section \ref{sec_ripples}.

\subsubsection{Brightness of the spine}

\begin{figure}
\includegraphics[width=0.5\textwidth]{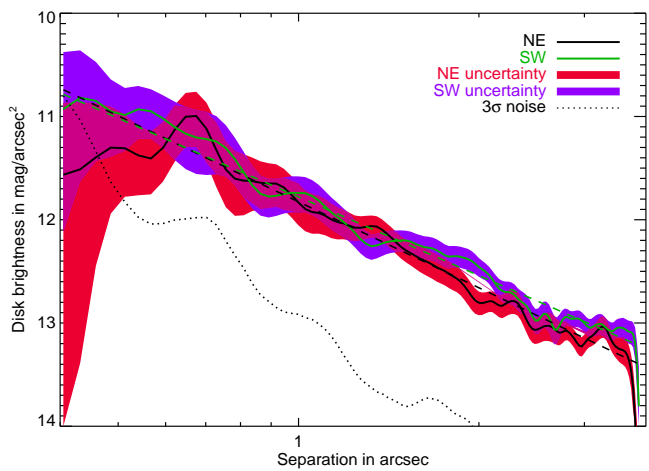}
\caption{Surface brightness of the spine, measured on the cADI-disc image from the fit of a Lorentzian profile. The red and purple shaded areas show the $3\sigma$ error in the amplitude of the Lorentzian function fitted to each vertical profile.}
\label{fig_disc_brightness}
\end{figure}

The surface brightness of the spine as a function of the separation is plotted in Fig. \ref{fig_disc_brightness}. The noise is evaluated as the standard deviation measured azimuthally in the regions where the disc is not detected. The two extensions of the disc show no overall brightness asymmetry within error bars inside $1.5\arcsec$. There is one localized asymmetry slightly inside $2\arcsec$ : the SW extension is locally 0.5 magnitude brighter than its NE counterpart. A localized brightness asymmetry was also detected in mid-infrared images of the disc by \citet{Telesco2005} at wavelengths beyond $\unit{8}{\micro\meter}$, but at a larger separation, namely $2.7\arcsec$. In the optical, the SW extension is also slightly brighter than the NE extension between 50 and 100 AU or  $2.5\arcsec$ and $5\arcsec$ \citep{Golimowski2006}.

The surface brightness profile appears smooth between $0.5\arcsec$ and $3.8\arcsec$, compatible with a single power-law dependance with the separation. The inflection seen in the optical at $2\arcsec$ by \citet{Golimowski2006} is not detected in our data. A linear regression to the NE and SW extensions between $0.5\arcsec$ and $3.7\arcsec$ yields slopes of $-2.77\pm0.18$ and $-2.57\pm0.16$.  The linear fit is shown as a dotted line in Fig. \ref{fig_disc_brightness}. The steeper slope of the NE extension explains why it appears slightly fainter beyond $1.5\arcsec$. The error bar shown in Fig. \ref{fig_disc_brightness} only includes the measurement error. Although the image was corrected for self-subtraction by applying the iteration technique described in section \ref{sec_bias_correction}, significant self-subtraction could still occur inside $0.8\arcsec$, therefore, the brightness inside $0.8\arcsec$ should be considered as a lower bound. To overcome this difficulty, we performed forward modelling with an innovative approach.

\section{Forward modelling}
\label{sec_interpretation}
\subsection{Modelling philosophy}

We used scattered light disc models to interpret the observed features and to disentangle ADI artifacts from real features. The disc models were generated with the GRaTeR code \citep{Augereau1999}. For each of the seven individual cubes of frames, the disc model is rotated to the appropriate parallactic angles of the initial frames and then subtracted. The resulting cubes are re-reduced using the same PCA algorithm as described previously. The seven reduced images are then combined to obtain one single disc-subtracted image. These steps are repeated iteratively by varying the free parameters of the disc model until a merit function is minimized. The minimization algorithm is a downhill simplex method or amoeba. For each minimization, three different sets of initial conditions were explored, all representing  physically acceptable conditions to reduce the risk of finding local minima. We found that all sets agreed within less than $5\%$. The merit function is a reduced chi squared computed in the part of the image where the disc is detected\footnote{This region includes 11363 pixels and the number of free parameters for the model is 4, which brings the number of degrees of freedom for the reduced chi squared to 11359.}. \citet{Thalmann2014} used this forward modelling approach as well to study the disc of LkCa\ 15. 

\subsection{Assumptions}
\label{sec_single_pop} 

 We neglected thermal emission over scattering by the dust particles. We discuss this assumption further in section \ref{sec_discussion} where we show that, even if the thermal emission might not be negligible depending on the exact grain properties (especially their composition), the thermal emission cannot fully dominate the scattered light above $0.5\arcsec$ and the main conclusion concerning the dust distribution will be unchanged. To limit the parameter space, we assumed an azimuthally symmetric distribution of optically thin dust with constant effective scattering cross-section. We used a Henyey-Greenstein phase function, whose single parameter $g$ is not computed from theoretical dust optical properties but is adjusted to the data. We did not attempt to model the warp in a first approach but used a single population model similar to the single dust population presented in \citet{Ahmic2009}.

We followed the work of \citet{Augereau2001}, called A01 hereafter, who derived the surface density of the parent belt (PB) of planetesimals in the disc from dynamical modelling. Their model locates the PB between 50 and 120AU and can explain both the surface brightness of the disc in scattered light and its infrared emission. It can also reproduce the recent submillimetric observations at $870\mu m$ obtained with ALMA with a resolution of $\sim$$0.6\arcsec$  \citep{Dent2014} very well. We therefore assumed a dust distribution matching that of the PB within this range. Beyond a radius $r_{out}=115AU$, we extrapolated the dust surface density with power laws to account for the grains that are blown away by radiation pressure (but still bound to the star) and that populate the regions outside the outer edge of the parent belt. \citet{Pantin1997} argues that the dust density should decrease for distances larger than 117AU following a power law  with an index within the range [-1.5;-2.3] according to visible observations. This is in agreement with theoretical expectations (e.g. \citealt{Thebault2008}). We therefore tested three different laws of indices $p_{out}=-1.5, -1.7$ and $-2.3$. Little is known about the presence of dust within 50AU. Resolved mid-infrared images by \citet{Pantin1997} and modelling of the spectral energy density by \citet{Li1998} suggest that most of the dust lies beyond 30AU but do not exclude an inner dust population around 10AU. This is supported by the work of \citet{Okamoto2004} who revealed what may be several planetesimal belts of amorphous silicate grains. In this context, we consider three different scenarios for the dust distribution within 50AU. In the first scenario, little dust lies inside 50AU and the surface density follows that of the PB. In the second and third scenario, the dust surface density follows a power law of index $p_{in}=0.5$ and $1.5$ respectively.

\begin{figure}
\includegraphics[width=0.5\textwidth]{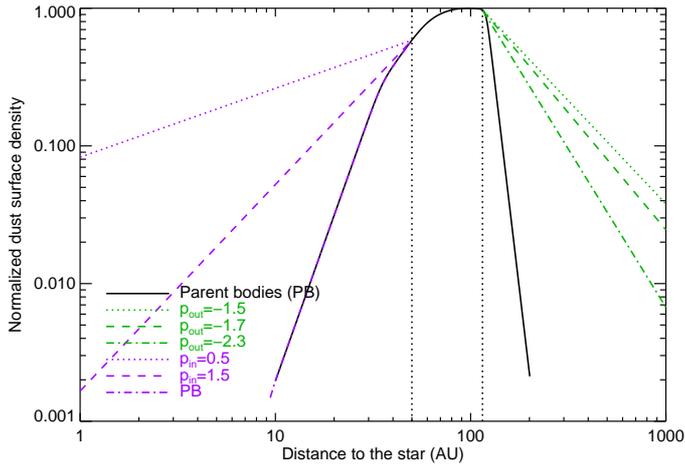}
\caption{Description of the dust surface density used in the nine different models, based on the surface density derived by \citet{Augereau2001} for the parent bodies (PB, black line). Each of the nine models is made of three parts delimited by the two vertical dotted lines at 50AU and 115AU: an inner region (purple curves) following either a power law or the PB distribution, a common intermediate region following the PB distribution, and an outer region following a power law (green curves).}
\label{fig_surf_density_augereau}
\end{figure}

All in all, we ended up with $3\times 3=9$ models for the dust surface density profile. They are summarized in Fig. \ref{fig_surf_density_augereau}. We had to further assume a vertical distribution $Z(r,z)$ for the dust to convert our surface density into volume density and compute the scattered light images. To do so, we followed \citet{Ahmic2009} who assumed that the dust vertical volumetric density follows an exponential profile with an exponent $\gamma = 0.5$, a scale height $\xi_0$ at the radius $r_{out}$, and a flaring exponent $\beta=1.5$ so that $Z(r,z)=\text{exp} \left[ -  \left( \frac{z}{\xi_0 (r/r_{out})^\beta} \right)^\gamma \right] $. 

We observed one degeneracy in the remaining free parameters of the disc. The inclination of the disc $i$ and the anisotropic scattering parameter $g$ are degenerate. More inclined discs require a smaller $g$ parameter to explain the offset and bow of the spine.  For this reason, we considered explicitly distinct inclination values $i$ of $89^{\circ}$, $88^{\circ}$, $87^{\circ}$, $86^{\circ}$, and $85^{\circ}$. We did not consider an edge-on disc  ($i=90^\circ$) because the offset of the spine can only be reproduced assuming anisotropic scattering combined with a slight inclination of the disc. Likewise, we did not explore an inclination below $i=85^\circ$ because it produces a disc significantly thicker than observed.  
The remaining free parameters for the model are therefore the PA, $\xi_0$, $|g|$, and a flux scaling factor.

\subsection{Results}
\label{sec_result_single_comp}
\begin{figure}
\centering                          
\includegraphics[width=0.5\textwidth]{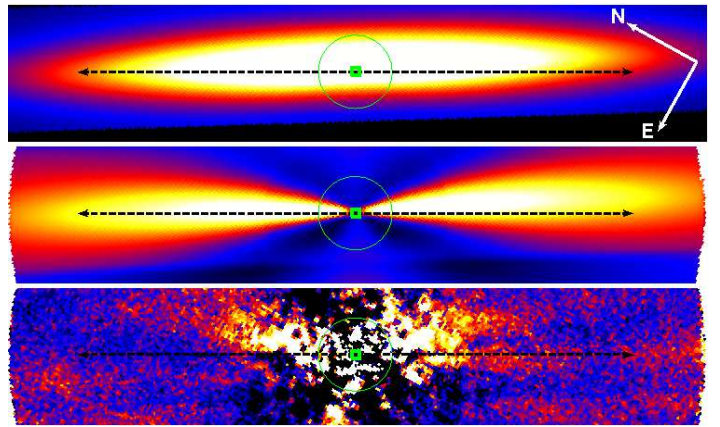}
\caption{From top to bottom: Initial disc model, disc after PCA reduction without noise, residuals remaining after the disc model was subtracted from the data cubes and the cubes were re-reduced. The colour scale is linear and two times larger for the initial model than for the other two images. The green circles have a radius of $0.4\arcsec$ and the black arrows indicate the apparent midplane (PA=$29^\circ$) with a total length of $6\arcsec$.}
\label{fig_disc_subtraction}
\end{figure}

We show in Table \ref{tab_result_minimization} the best parameters corresponding to the most relevant cases in the $(i,p_{in},p_{out})$ exploration space. They are relatively robust to the different scenarios investigated. The values of the best PA all lie between $30.6^\circ$ and $30.9^\circ$, for instance, and for a given inclination, we observe little dispersion in the parameters $|g|$ and $\xi_0$.
Overall, the best $\chi^2$ is obtained with an inclination $i$ of $86^\circ$, and the three best $\chi^2$ are obtained with an index $p_{in}$ of $1.5$, which is an indication that the dust surface density is not as steep as that of the PB within 50AU. The best model is shown in Fig. \ref{fig_disc_subtraction} (top image), while the middle image shows the disc model after PCA-reduction without noise to illustrate the biases of ADI. We also show the residuals after subtraction of the model from the data (bottom image). Disc residuals are still present in this model-subtracted image inside $1\arcsec$, mainly above the apparent midplane, and this will be discussed in the next section.

For all disc models with an inclination $i=86^\circ$, we observe that an inner dust distribution following that of the PB systematically yields the worst $\chi^2$, and a value of $p_{in}=1.5$ always obtains the best figures of merit. The different values of $p_{out}$ yield very similar results. For a higher inclination of $88^\circ$, models are more anisotropic and with larger scale heights. This corresponds to the degeneracy between i and g mentioned earlier and to the fact that a vertically thicker disc is needed when the inclination is closer to edge-on. For such an inclination, we can reasonably rule out the less steep inner dust distribution $p_{in}=0.5$ because it systematically leads to a higher $\chi^2$.

\definecolor{Gray}{gray}{0.9}
\begin{table}
\caption{Parameters corresponding to the best disc models for different fixed values of the inclination $i$ and power law indices $p_{in}$ and $p_{out}$. The other parameters are free.}             
\label{tab_result_minimization}      
\centering                          
\begin{tabular}{ccccccc}       
\hline     \hline     
  $i$ & $p_{in}$ & $p_{out} $ & PA  & |g| & $\xi_0$\tablefootmark{a} (AU) & $\chi^2$ \\    
\hline                        
$89^\circ$ & 1.5 & -1.7  &  $30.7^\circ$  &  0.57  & 7.9 & 2.62 \\       
\hline                        
$88^\circ$ & 0.5 & -1.5  &  $30.7^\circ$  &  0.49  & 8.6 & 1.49 \\       
$88^\circ$ & 1.5 & -1.5  &  $30.7^\circ$  &  0.45  & 8.9 & 1.35 \\       
$88^\circ$ & PB & -1.5  &  $30.7^\circ$  &  0.47  & 8.7 & 1.38 \\       

$88^\circ$ & 0.5 & -1.7  &  $30.6^\circ$  &  0.49  & 8.7 & 1.50 \\       
$88^\circ$ & 1.5 & -1.7  &  $30.8^\circ$  &  0.44  & 8.6 & 1.35 \\       
$88^\circ$ & PB & -1.7  &  $30.8^\circ$  &  0.47  & 8.7 & 1.38 \\       

$88^\circ$ & 0.5 & -2.3  &  $30.6^\circ$  &  0.50  & 8.5 & 1.52 \\       
$88^\circ$ & 1.5 & -2.3  &  $30.7^\circ$  &  0.44  & 8.6 & 1.36 \\       
$88^\circ$ & PB & -2.3  &  $30.8^\circ$  &  0.47  & 8.6 & 1.39 \\       

\hline                        
$87^\circ$ & 1.5 & -1.7  &  $30.8^\circ$  &  0.42  & 8.7 & 1.31 \\       
\hline                        
$86^\circ$ & 0.5 & -1.5  &  $30.7^\circ$  &  0.33  & 6.0 & 1.37 \\       
$86^\circ$ & 1.5 & -1.5  &  $30.9^\circ$  &  0.31  & 4.9 & 1.31 \\       
$86^\circ$ & PB & -1.5  &  $30.8^\circ$  &  0.30  & 5.1 & 1.44 \\       

$86^\circ$ & 0.5 & -1.7  &  $30.7^\circ$  &  0.36  & 6.3 & 1.37 \\       
$86^\circ$ & 1.5 & -1.7  &  $30.9^\circ$  &  0.30  & 5.3 & 1.31 \\       
$86^\circ$ & PB & -1.7  &  $30.8^\circ$  &  0.30  & 5.3 & 1.44 \\       

$86^\circ$ & 0.5 & -2.3  &  $30.7^\circ$  &  0.36  & 6.0 & 1.36 \\       
\rowcolor{Gray}
$86^\circ$ & 1.5 & -2.3  &  $30.8^\circ$  &  0.36  & 5.1 & 1.30 \\       
$86^\circ$ & PB & -2.3  &  $30.8^\circ$  &  0.36  & 4.9 & 1.43 \\       
\hline                        
$85^\circ$ & 1.5 & -1.7  &  $30.8^\circ$  &  0.22  & 3.7 & 1.33 \\       
\end{tabular}
\tablefoot{
\tablefoottext{a}{\small{Scale height at 115 AU}}
}
\end{table}

We conclude from this analysis that the PA of the disc as measured within the $3.8\arcsec$ field of view lies between $30.9^\circ$ and $30.6^\circ$. This is in agreement with the value of the PA measured from our single-lorentzian fit: the PA of the disc as measured inside $3.8\arcsec$ is greater than that measured at larger distances where the warp component is not detected. 

The disc scatters light anisotropically with a value of the Henyey-Greenstein parameter $g$ dependent on the inclination of the disc. For a disc inclined by $89^\circ$, a relatively strong anisotropic scattering with $|g|=0.6$ is necessary to reproduce the bow of the disc. There is a moderate anisotropy ($|g|=0.4$ to $0.5$) for a  disc inclined by $88^\circ$, and only a slight anisotropy ($|g|=0.3$ to $0.4$) for an inclination of $86^\circ$. If this anisotropy is because of the enhanced forward-scattering efficiency of the dust, as is often claimed \citep{Golimowski2006}, then the NW side is the closest to the Earth. These g values are compatible with those derived in the near-infrared for other debris discs, for instance $|g|=0.3\pm0.03$ for HD\ 181327 \citep{Schneider2006}, and \citet{Rodigas2012} could also explain the bow of HD\ 15115 detected at $L^\prime$ with an anisotropic scattering parametrized by $|g|=0.5$.

\section{Discussion}
\label{sec_discussion}

\subsection{Possible scenarios to explain the missing flux inside $1\arcsecond$}
\label{sec_validation}

\begin{figure}
\includegraphics[width=0.5\textwidth]{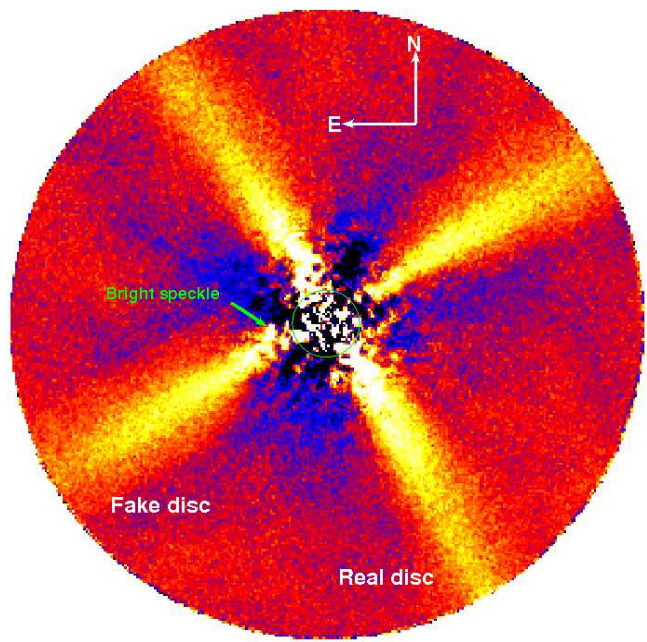}
\caption{Image obtained after injection of the best disc model at $90^\circ$ from the real disc and reducing the data again. The field of view is $3.8\arcsec$ and the real disc extends from NE to SW.}
\label{fig_fake_disc_90_calibration}
\end{figure}

To test further the agreement of our models with the data, we injected a fake disc (our best-fit model for $i=86^\circ$, $p_{in}=1.5$ and $p_{out}=-2.3$) at a PA $90^\circ$ away from the disc in the data cubes. The result is shown in Fig. \ref{fig_fake_disc_90_calibration}. Visually, the model looks very similar to the observed disc beyond $1\arcsec$ but appears too faint inside that region, as already seen in the model-subtracted image of Fig. \ref{fig_disc_subtraction}. Quantitatively, the reduced fake disc is not detected inside $0.5\arcsec$ and lacks about $0.3\  \text{mag/arcsec}^2$ at $0.8\arcsec$. 

\begin{figure}
\includegraphics[width=0.5\textwidth]{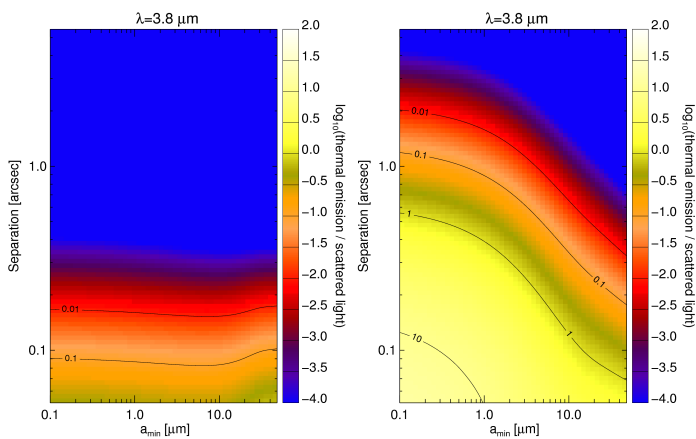}
\caption{Flux ratio between the dust thermal emission and the scattered light integrated along the line of sight assuming our best-fit dust distribution and either porous astronomical silicates (left) or the more complex dust aggregates described in the text (right).}
\label{fig_ratio_th_scat}
\end{figure}

We discuss three possibilities to explain the excess emission within $1\arcsec$ compared to our best scattered light model:
\begin{enumerate}
\item There is more dust inside $20AU$ than predicted by our model. This warm dust would therefore enhance the scattered light from the star and could also contribute to thermal emission at $L^\prime$. \citet{Wahhaj2003} estimate from thermal imaging at $\unit{17.9}{\micro\meter}$ that there is indeed an inner ring at 14AU. \citet{Pantin1997} showed that mid-infrared observations are compatible with a dust depletion around 20AU, but the surface density could increase again inside 10AU (see the solid red curve in Fig. \ref{fig_comp_pantin} described later in section \ref{sec_comparison_pantin}). \citet{Augereau2001} also proposed a population of hot grains within 20 AU to explain the $12 \mu m$ flux. Last, interferometric measurements showed that the star exhibits a near-infrared excess at H and Ks \citep{diFolco2004,Defrere2012} that cannot be entirely explained by scattered light from the edge-on disc within the interferometric field of view. All these observations are supported by the fact that none of our models can explain the strong brightness of the disc within $1\arcsec$. To investigate this possibility, we used GRaTeR to compute the amount of scattering and thermal emission occurring along the line of sight for our best-fit dust distribution model, assuming two different types of grain populations: porous astronomical silicates \citep{Draine2003} or more realistic aggregates\footnote{The composition is that of \citet{Augereau2001} with volume ratios of carbonaceous material, astronomical silicates, water ice, and vacuum set to $2:1:6.2:82.5$ respectively. References for the optical constants and how the optical properties are computed can be found in \citet{Lebreton2012}.} made of a silicate core coated by an organic refractory mantle, with water ice partly filling the holes created by the porosity, if the temperature is less than the water sublimation temperature \citep{Augereau1999}. In both cases, the grain size distribution $\frac{dn}{da}$ follows the traditional $a^{-3.5}$ power-law decrease resulting from a collisional cascade with a minimum size $a_{min}$. The result is shown in Fig. \ref{fig_ratio_th_scat}. It confirms that porous silicates cannot explain the missing flux because scattering predominates by a factor over 100 beyond $0.2\arcsec$. For the population of more complex aggregates, we cannot exclude a contribution of thermal emission dominant up to $0.5\arcsec$ and still significant between $0.5\arcsec$ and $1\arcsec$, especially if the grain population is micronic or sub-micronic, which is expected from the population in the inner 20AU \citep{Augereau2001}. At $0.8\arcsec$, a ratio between thermal emission and scattered light of $0.3$ is enough to explain the $0.3mag$ missing flux, which is obtained for a grain population of minimum size below $0.8\mu m$ (Fig. \ref{fig_ratio_th_scat}).
\item The Henyey-Greenstein scattering phase function we used does not reproduce the scattering properties of the dust grains well. The theoretical phase function calculated in the frame of the Mie theory, assuming hard silicate spheres of a few microns, predicts a much more peaked forward-scattering behaviour (see for instance \citealt{Mulders2013}). If we assume the flux within $1\arcsec$ is mostly due to forward scattering events occurring around 100AU, then the scattering phase angles are below $11^\circ$, thus well within the range where the Mie theory predicts a strong peak, and this could explain our missing flux. This may also explain why the excess emission in Fig. \ref{fig_disc_subtraction} (bottom) lies above the disc midplane. The scattering phase function is very sensitive to the shape and structure of the grains, however, and it remains to be proven that such theoretical phase functions are realistic for porous fluffy grains that can significantly deviate from a spherical shape \citep{Augereau1999,Schneider2006}.
\item Our forward modelling approach cannot constrain the dust distribution at short separation. The optimization region where the $\chi^2$ is minimized starts at  $0.4\arcsec$ or 7.7AU, so any change in the disc brightness or morphology within $0.4\arcsec$ does not affect the  $\chi^2$. Pixels at  $0.4\arcsec$ also have a smaller weight in the  $\chi^2$ because the noise is higher at that separation, which could introduce a bias towards more external regions. This explanation alone cannot explain the missing flux observed up to $1\arcsec$. An additional explanation might be the fact that the real and fake disc interfere during the star subtraction process, because the field rotation is close to $90^\circ$ in some data sets. This would reinforce self-subtraction in the inner regions but both the fake and real disc would be similarly self-subtracted, whereas we do not notice a flux change in the real disc between Fig. \ref{fig_fake_disc_90_calibration} and Fig. \ref{fig_disc_cadi_pca}.
\end{enumerate}
Those three scenarios  do not exclude each other. Based on the observational evidence mentioned above, especially the mid-infrared imaging, we favor the first scenario, but a combination of the three is very likely.

%

\subsection{The reality of the ripples}
\label{sec_ripples}

\begin{figure}
\includegraphics[width=0.5\textwidth]{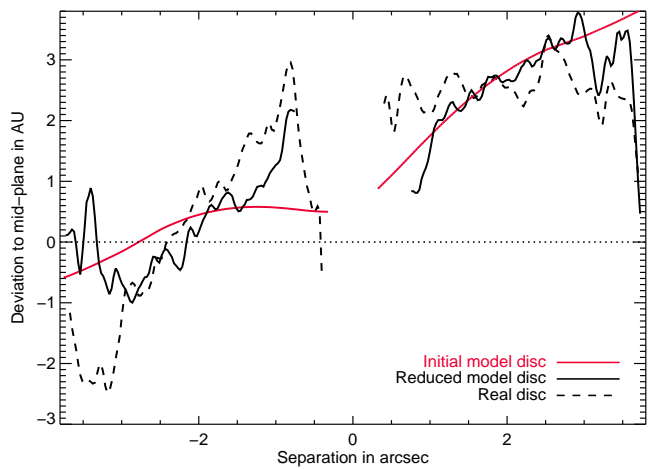}
\caption{Departure of the spine with respect to the apparent midplane as measured on the initial disc model (top right image of Fig. \ref{fig_disc_subtraction}), on the fake disc injected at $90^\circ$ (Fig. \ref{fig_fake_disc_90_calibration}) and on the real disc. NE is to the left for the real disc and SE is to the left for the fake disc.}
\label{fig_disc_center_calibration}
\end{figure}

We measured the departure of the spine with respect to the apparent midplane on the fake disc with the same technique as that applied to the real disc, and show the result in Fig. \ref{fig_disc_center_calibration} (black solid line). On the same graph, we plot the measurement as performed on the real disc (dashed line), and performed on the initial model (red line). 

First, the measurement performed on the fake disc injected at $90^\circ$ also shows ripples with the same amplitude as those measured on the real disc. Therefore, the ripples detected longwards $1\arcsec$ on each extension in Fig. \ref{fig_spine_position} are likely to be caused by the remaining noise. Inside $1\arcsec$, the fake disc spine measurements vary significantly  before and after PCA reduction. In particular, we also detect a ripple on the left side at about $0.8\arcsec$ with more than half the amplitude of that detected on the real disc. This is because of a very bright speckle located at a PA of about $110^\circ$ (see Fig. \ref{fig_fake_disc_90_calibration}, green arrow, also visible in Fig. \ref{fig_disc_cadi_pca}), very close to the fake disc PA of $120.8^\circ$. This speckle clearly tends to shift the spine of the fake disc upwards, as visible in Fig. \ref{fig_fake_disc_90_calibration}. We should emphasize that at this separation the reduced fake disc is $0.3$ magnitude fainter than the real disc, so the measurement is more sensitive to the underlying speckle noise.
Therefore, the current SNR does not allow us to rule out the fact that the deviation of the spine detected at $0.8\arcsec$ might be due to the residual noise. This deviation is, however, twice as large as that induced by remaining speckles and is detected on three of the seven individual images (those with the best SNR), which may not be a coincidence. 

Second, we note some significant variations between the departure of the spine as measured on the initial disc model and on the real data. While the value is, on average, close to 0 and 2AU for the NE and SW side, the slope is too steep for the SW side and not steep enough for the NE side. We will describe this behaviour in more detail in section \ref{sec_double_pop} with a two-component disc model.

\subsection{Comparison with mid-infrared observations}
\label{sec_comparison_pantin}

\begin{figure}
\includegraphics[width=0.5\textwidth]{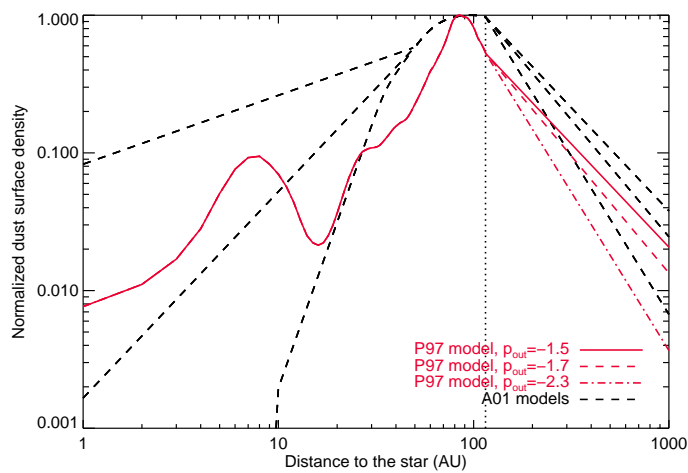}
\caption{Comparison between the P97 and A01 models. The vertical dotted line indicates the separation of 115AU beyond which the dust surface density inverted from the $12\mu m$ observations was interpolated with power laws.}
\label{fig_comp_pantin}
\end{figure}

In mid-infrared around $10 \mu m$, thermal emission by dust is the dominant process over scattering in the inner 100AU and observations are not limited by the contrast to the star while they still provide a resolution of 5AU \citep{Pantin1997,Okamoto2004,Telesco2005}. It is therefore relevant to compare the dust distribution derived from our images to that derived from mid-infrared observations. Using observations at $12 \mu m$ from the TIMMI camera, \citet{Pantin1997} proposed a model of the dust distribution in the inner 100AU, by inverting the integral equation relating the spatial flux to the dust surface density. Their choice of grain composition and size distribution led to the dust surface density displayed in Fig. \ref{fig_comp_pantin} (plain red line), later referred to as P97. This profile corresponds to the average between NE and SW surface density presented in Fig. 5 of \citet{Pantin1997}, after correcting for the revised distance of $\beta$ Pictoris of 19.3pc. 
Again, building a scattered light image using this distribution requires extrapolating the surface density beyond 100AU, and we followed the same approach as described previously with different power laws of indices $p_{out}=-1.5, -1.7$, and $-2.3$.
The comparison between the P97 and A01 models is shown in Fig. \ref{fig_comp_pantin}. The dust surface density from P97 models predicts a narrower peak density and varies significantly from our models within the inner 20AU, predicting more dust within 10AU than our (A01,$p_{in}=0.5$) and (A01, $p_{in}=1.5$) models. We chose the same dust vertical distribution as our models, e.g. an exponential profile of index $\gamma=0.5$, a flaring exponent $\beta=1.5$, and a scale height $\xi_0$ at $r_{out}=115AU$, and run our minimization algorithm to subtract the best parametric model from our data. The results are shown in Table \ref{tab_result_minimization_pantin}.  The P97 model does not explain the data better than our models with a $\chi^2$ degraded by $3.6\%$ and $14\%$ on average for an inclination of $88^{\circ}$ and $86^{\circ}$, and the disc-subtracted image still displays some disc residuals within $1\arcsec$. We can conclude that the A01 models better reproduce our observations beyond $0.5\arcsec$ or 10AU. Moreover, our estimate of the parameters PA, $\xi_0$ and $|g|$ is robust to the underlying density distribution because we find similar values as in Table \ref{tab_result_minimization}.  Little can be said inside $0.4\arcsec$ on the reality of the rise in dust surface density at about $8AU$ in P97 (see Fig. \ref{fig_comp_pantin}) for two reasons: we are probably not dominated by scattering at this separation, especially if such a warm dust population is present, and the exact scattering properties of the dust at small phase angles are still uncertain.

\begin{table}
\caption{Parameters corresponding to the best disc models following P97, for different fixed values of the inclination $i$ and power-law index $p_{out}$.}      \label{tab_result_minimization_pantin}      
\centering                          
\begin{tabular}{ cccccc }        
\hline     \hline     
  $i$  & $p_{out}$ & PA  & |g| & $\xi_0$(AU)& $\chi^2$ \\    
\hline                        
$88^\circ$  & -1.5 &  $31.2^\circ$  &  0.63  & 8.2 & 1.48 \\       
$88^\circ$  & -1.7 &  $31.2^\circ$  &  0.63  & 8.2 & 1.45 \\       
$88^\circ$  & -2.3 &  $31.2^\circ$  &  0.61  & 8.2 & 1.47 \\       
\hline                        
$86^\circ$  & -1.5 &  $31.0^\circ$  &  0.27  & 4.7 & 1.57 \\       
$86^\circ$  & -1.7 &  $31.0^\circ$  &  0.27  & 4.4 & 1.57 \\       
$86^\circ$  & -2.3 &  $30.8^\circ$  &  0.26  & 4.6 & 1.56 \\       
\end{tabular}
\end{table}

\subsection{Disc model with an inner and outer component}
\label{sec_double_pop}

\begin{figure}
\includegraphics[width=0.5\textwidth]{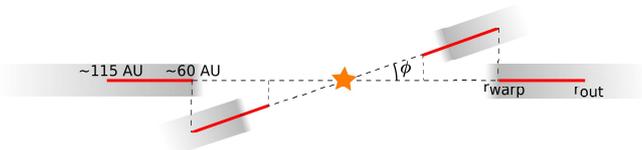}
\caption{Schematic of the disc model including a main and warped component. For clarity, the disc is shown edge-on, on a vertical cut. The red line indicates the region where most of the dust would lie in the A01, $p_{in}=1.5$ model.}
\label{fig_schematic}
\end{figure}

Among the various disc parameters compatible with the data, it clearly appears that the PA differs significantly from the value measured in scattered light at larger separations beyond $5\arcsec$. In an attempt to better understand this discrepancy and take advantage of our knowledge of the disc from previous modelling, we used a more sophisticated model to take the warped component of the disc into account. We therefore introduced a two-component model, as sketched in Fig. \ref{fig_schematic}, with a main outer disc and a warped inner disc. 
We have too limited a field of view and lack a signal from the main outer disc to constrain the surface density of the outer component and lead the same forward modelling approach as for the single-component model presented in section \ref{sec_single_pop}. This will be the purpose of a future analysis using multiple-wavelength data and scattered-light images with a larger field of view. Instead, we performed a qualitative analysis to understand if this more complex model can better explain the position of the spine. 
We assumed that the warped inner disc has a surface density identical to our A01, $p_{in}=1.5$ model until the separation $r_{warp}$, and extrapolated with a power law of index $p_{out}=-1.7$ beyond $r_{warp}$. The main outer disc surface density is 0 inside $r_{warp}$ and follows our A01, $p_{in}=1.5$, $p_{out}=-2.3$ model  beyond. 

We further assumed that each of the two components can be described with the same parameters $\beta=1.5$, $\gamma=0.5$, inclination $i=86^\circ$, anisotropy coefficient $|g|=0.36$ and scale height $\xi_0=5.1AU$ at 115AU, taken from the best model of Table \ref{tab_result_minimization}. The specific parameters of each components are:
\begin{itemize}
\item for the outer main disc, the inner radius  $r_{warp}$ and the PA of $29^\circ$
\item for the inner warped disc, the outer radius $r_{warp}$, and the PA offset by $\phi=4^\circ$ with respect to the outer component
 \end{itemize}

\begin{figure}
\includegraphics[width=0.5\textwidth]{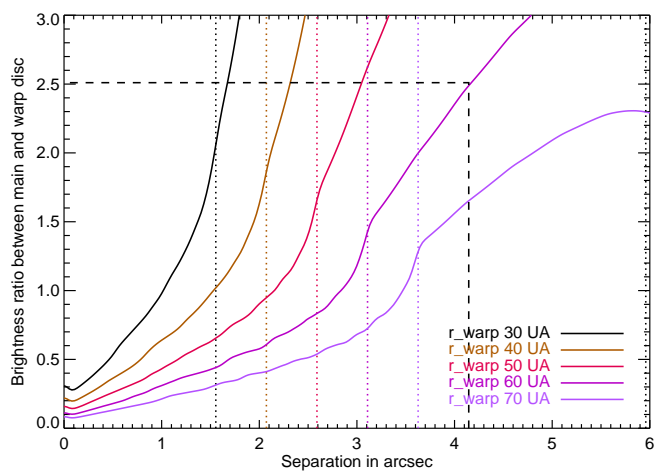}
\caption{Brightness ratio between the main outer disc and the warped inner disc for different values of $r_{warp}$. The dashed line indicates the ratio of 2.5 measured by \citet{Golimowski2006} at 80AU or $4.15\arcsec$. The vertical dotted lines indicate the different values of $r_{warp}$.}
\label{fig_disc_calibration_radius_warp}
\end{figure}

In this model, the only free parameters are now the radius $r_{warp}$ and the flux scaling factor. To constrain $r_{warp}$, we used the flux ratio between the main and the warped disc as measured from the two-component fit of vertical profiles by \citet{Golimowski2006} on the images from the ACS instrument aboard the Hubble space telescope. A ratio of 2.5 was measured at a separation of $80 AU$. If we assume that this ratio does not depend on the wavelength, this translates into a value of $r_{warp}$ of about 60AU as shown in Fig. \ref{fig_disc_calibration_radius_warp}. Interestingly, Fig. \ref{fig_disc_calibration_radius_warp} shows us that the main source of emission seen within $2.8\arcsec$ comes from light scattered by the warped disc rather than by the main disc. The brightness ratio between the two components of the disc is indeed smaller than 1 inside $2.8\arcsec$.

\begin{figure}
\includegraphics[width=0.5\textwidth]{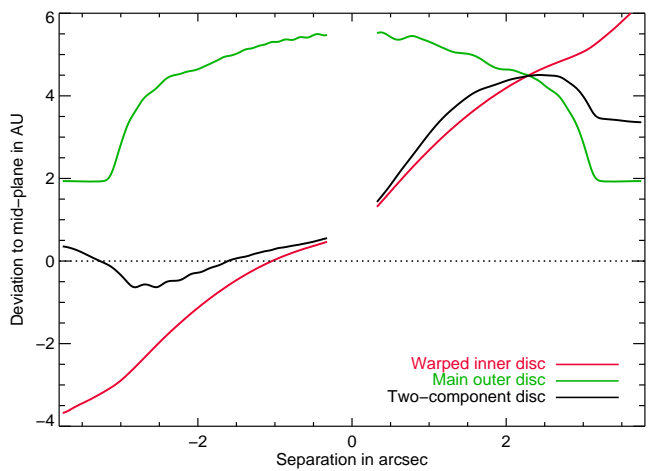}
\caption{Departure of the spine with respect to the apparent midplane as measured on a disc model presenting an outer disc aligned with a PA of $29^\circ$, plus an inner warped component extending up to $60$AU. The black curve reproduces the behaviour of the measured departure of the spine well, as shown in Fig, \ref{fig_disc_center}.}
\label{fig_disc_calibration_warp}
\end{figure}

Using this model, we then performed our single-lorentzian fit again after derotating the image by the PA of the main disc, i.e. $29^\circ$ (Fig. \ref{fig_disc_calibration_warp}). We also display the fit done on each individual components of the model. The black curve shows the same trend as our measured spine, namely an overall offset above the apparent midplane inside $2\arcsecond$ with a larger offset on the NE than on the SW extension. We clearly see that the spine position of the double-component model lies in between that of the single-component models and tends to align with that of the main outer component beyond $3\arcsec$. Within $3\arcsecond$, the departure of the spine with respect to the apparent midplane is steeper on the NE side than on the SW side, as seen in our data in Fig, \ref{fig_disc_center}.

This model also produces a subtle brightness asymmetry of $0.1$ mag/arcsec$^2$ between the NE and SW extension inside $2\arcsecond$. Indeed, the spine position of the main and warped components (green and red curves in Fig. \ref{fig_disc_calibration_warp}) being closer to each other on the SW side, the brightness of the summed contribution is slightly larger on the SW side than on the NE side. This difference is compatible with our measurements and can explain the slight asymmetry noted between $1.5\arcsecond$ and $2\arcsecond$ in Fig. \ref{fig_disc_brightness}. 

All in all, this two-component model can explain the offset and bow of the disc. The presence of the warped component creates an asymmetry with a departure of the spine greater on the NE side than on the SW side, as observed in the data. 

\section{Conclusion}
\label{sec_conclusion}

\begin{enumerate}
\item We have obtained $L^\prime$-band images of $\beta$ Pictoris that reveal the disc in its innermost regions inside $1\arcsec$ dominated by light scattered from the warped inner component. The surface brightness distribution shows no overall asymmetry between the two extensions of the disc within error bars. The spine is offset and bowed due to anisotropic scattering combined with a small inclination of the disc estimated around $86^\circ$. A large ripple is detected in the spine position on the NE side at $0.8\arcsec$ with a confidence level of $2\sigma$, although we cannot rule out that it comes from the residual speckle noise. If real, this deviation of the spine might be the sign of a gravitational perturbation. We do not detect any obvious point-source within the disc after subtracting the signal from $\beta$ Pictoris b, but two aspects should be highlighted. First, the brightness of the disc hinders the detection of faint point-sources. Second, as a result of our strategy to combine multi-epoch data, any planet on an orbit smaller than that of $\beta$ Pictoris b would smear an arc on the combined image of the seven epochs and its flux would be diluted over more than one resolution element. A paper dedicated to the detection limits on potential planetary companions located in the disc midplane is in preparation. The paper will make use of  these combined data sets to go deeper than in \citet{Absil2013} and will be complemented by radial velocity constraints.
\item The observations are compatible with the presence of dust within 10AU from the star, although degeneracies in the modelling prevents us from accurately constraining the inner dust distribution. Our single-component disc model explains the overall morphology of the disc well. We favour a best model inclined by $86^\circ$ with an anisotropic scattering coefficient $|g|$ of $0.36$. However, this model cannot explain the strong brightness of the innermost parts of the disc inside $1\arcsec$. Different scenarios can be proposed to explain why our models fail to predict this strong brightness, the most likely being either that thermal emission starts to be significant {inside} $0.5\arcsec$ or that enhanced forward scattering occurs for small phase angles. Additional observations in scattered light and at shorter wavelengths insensitive to thermal emission, but with a sufficient resolution and contrast to probe the inner $0.5\arcsec$ are necessary to disentangle those two scenarios. 
\item An investigation of a two-component disc model, mimicking a warped disc extending up to 60AU and offset by $4^\circ$ with respect to a main outer disc extending from 60 to 115AU shows that within $2.8\arcsec$ the scattered light mainly comes from the warped inner disc. The measured position of the spine is well reproduced by such a model, especially the bow and offset. A multi-wavelength modelling with a larger field of view is now needed to confirm our understanding of the dust distribution and bring additional constraints on the parameters derived in this study. 
\end{enumerate}
 
 \begin{acknowledgements}
O.A. acknowledges financial support from a FNRS Research Associate during part of this work and J.M. acknowledges financial support from the ESO studentship program and the Labex OSUG 2020. We also would like to thank the referee Dr. A. Brandecker for his constructive comments that helped to significantly improve the quality of the article.
\end{acknowledgements}

\bibliography{biblio} 	

\end{document}